\documentclass[12pt]{article}
\usepackage[utf8]{inputenc}
\usepackage{amsmath}
\usepackage{amsfonts}
\usepackage{amssymb}
\usepackage{hyperref}
\usepackage{slashed}

\title{\bf Scale Invariance at the Edge}
\author{Gordon W. Semenoff\\~
\\Department of Physics and Astronomy, University of British Columbia,\\
6224 Agricultural Road, Vancouver, British Columbia, Canada V6T 1Z1}
\date{}

\begin{document} 
\maketitle

\begin{abstract}Some aspects of the theory of fermions living on three dimensional spacetime with a
flat co-dimension one boundary are discussed, particularly a case where the boundary condition preserves scale and translation invariance but violates the residual Lorentz and conformal symmetries.  This case is interesting because the Dirac equation has normalizable stationary edge states which must be taken into account in
the quantization. We show that a consequence of the edge states for quantization of the Dirac field is that
there are no states of the Dirac field theory which are simultaneously scale, C and P invariant even when these are good symmetries of the theory. The scale invariant states of the Dirac field contain either a nonzero scale covariant momentum density or a U(1) charge density concentrated near the edge.
\end{abstract}
\flushbottom

 \section{Introduction}

The quantization of the Dirac theory on a space with a boundary requires boundary conditions.   There is the obvious question, especially with fermions, as to whether the boundary conditions respect the symmetry of the theory.   For example, the free massless Dirac theory in the bulk of 3 dimensional Minkowski space-time is a (trivial example of a) conformal field theory with the full $SO(3,2)$ conformal symmetry group.  Consider the same problem on a half-space -- 3 dimensional Minkowski space with an infinite, flat, co-dimension one boundary which is itself a 2 dimensional Minkowski spacetime.  One would expect that the bulk $SO(3,2)$ invariance would be reduced the $SO(2,2)$ symmetry of the boundary.  This symmetry consists of translation in time and space along the boundary, a Lorentz transform along the direction of the boundary and scaling and a conformal transformation of the boundary.  

However, the existence of this residual symmetry can of course depend on the boundary condition. There turns out to be a boundary condition which is relevant to the study of some two dimensional Dirac materials which terminate on boundaries where the boundary condition breaks what would be some of the residual space-time symmetry.   In the example of graphene, perhaps the prototypical example of a Dirac material \cite{Semenoff:1984dq}\cite{CastroNeto:2007fxn}, the edge that we are referring to has a zig-zag configuration \cite{ak1}\cite{ak2}.  The appropriate boundary condition has less spacetime symmetry than is possible for a massless Dirac field \cite{Biswas:2022dkg}.  It preserves only the scale and translation symmetry of the edge.  It violates Lorentz and conformal symmetry.   On the other hand, even the massless non-interacting Dirac Hamiltonian has interesting behaviour there.   It exhibits a continuous set of bound states at the edge.   

Bound states in a scale invariant field theory are unusual.  They must have vanishing binding energy.  Indeed they appear in the middle of the spectrum, between he positive and negative energy eigenstates of the Dirac Hamiltonian. In the following we will examine how the scale symmetry as well as the discrete spacetime symmetries are represented when these bound states are present.  We will identify scale invariant states of the Dirac field theory.  Those states have interesting properties in that requiring scale invariance will break some of the other symmetries of the theory, either charge conjugation C or parity P,  which are perfectly good symmetries of the Hamiltonian and the boundary conditions but seem to be incompatible with scale invariance.  The result, for scale invariant states will be, either a C invariant,P violating state which has a one-point function with non-vanishing momentum density $<{\bf T}^{02}(x)>$ or a $P$ invariant, $C$ violating state with a  one-point function with non-vanishing U(1) charge density.

  \section{Dirac theory on a 3D half-space}
  
 Consider the massless Dirac equation 
\begin{align}\label{de}
 \slashed \partial \psi^{(A)}(x)=0~,~~A=1,2
 \end{align}
 for two complex two-component spinors in 2+1-dimensions, with ``valley'' labels $A=1,2$,  
 defined on the half-space $x^\mu=(x^0,x^1,x^2)$ with  $x^1>0$ and with the boundary conditions
   \begin{align}\label{bc}
 \lim_{x^1\to 0} \left( 1+ i\gamma^0\right)\psi^{(1)}(x)=0
 ~,~~ \lim_{x^1\to 0} \left( 1- i\gamma^0\right)\psi^{(2)}(x)=0
 \end{align}
 We are considering the case where there are two valleys, $A=1,2$, 
 which have opposite boundary conditions in order
 to preserve some discrete spacetime symmetries, on which we will elaborate shortly.  These turn out to
 describe the continuum limit of the tight-binding model of graphene on its hexagonal lattice with a zig-zag edge, where we have assumed that the electron is spinless. 
 
The boundary condition is needed so that the normal component of the current vanishes at
the boundary 
\begin{align}\label{bc1}
\lim_{x^1\to 0}~\bar\psi^{(A)}(x)\gamma^1 \psi^{(A)}(x)  =0~,~~A=1,2
\end{align}
This is needed so that the Hamiltonian is Hermitian. An alternative, Lorentz invariant boundary condition 
would be $  
 \lim_{x^1\to 0} \left( 1\pm \gamma^1\right)\psi^{(A)}(x)=0$, similar to the MIT bag model boundary condition. It actually preserves the Lorentz and conformal symmetry of the boundary.  However, it does not allow edge states.  In the following we will focus on the Dirac fermion problem with edge states, that is, with the boundary condition (\ref{bc}).
 
 \section{Boundary condition breaks Lorentz and Conformal symmetry}
 
 In the bulk of three dimensional Minkowski space, the space-time symmetries of the fermions are the full set of set of conformal transformations.  In the presence of a boundary, these are reduced to the spacetime symmetries of the boundary, which are generated by spacetime Noether charges which have the algebra
 \begin{align}
   &-i\left[P_\mu,~\psi^{(A)} (x)\right]=\partial_\mu \psi^{(A)} (x) ~,~~\mu=0,2 \\
 &-i\left[  M,~\psi^{(A)}  (x)\right]= \left( x_2\partial_ 0-x_0\partial_2+\frac{1}{2}\gamma^1\right) \psi^{(A)}  (x)&  \label{Lorentz}\\
&-i \left[  \Delta,~\psi^{(A)}  (x)\right]=  \left(x^\mu \partial_\mu +1\right)~ \psi^{(A)} (x)&  \\
&-i\left[   {K_\lambda},~\psi^{(A)} (x)\right]=  \bigg\{\left(x_\lambda x^\mu-\frac{x^2 }{2} \delta_\lambda^\mu\right)\partial_\mu + \frac{1}{2} \epsilon_{\lambda\rho\sigma} x^\rho\gamma^\sigma +x_\lambda\biggr\} \psi^{(A)} (x)
,~\lambda=0,2
 \label{Conformal}
\end{align} 
However, we can see that the spin parts of the Lorentz transformation in equation (\ref{Lorentz}) and of the conformal transformation in equation (\ref{Conformal}) are incompatible with the boundary conditions  in equation (\ref{bc}).
This further reduces the symmetries to translations and scale transformations generated by
$P^0$, $P^2$ and $\Delta$. 

\section{CPT}

Consider the explicit representation of  the $2\times2$  Dirac matrices matrices  
\begin{align}\label{dirac matrices}
\gamma^0=-i\sigma^3~,~\gamma^1=\sigma^1~,~\gamma^2=\sigma^2
\end{align}
where $\sigma^a$ are the Pauli matrices.  
With this choice of Dirac matrices, the discrete spacetime symmetries  C, P and T are the transformations 
  \begin{align}
 &       
{\rm C}:(\psi^{(1)}(x),\psi^{(2)}(x))~\to~(\gamma^1\psi^{*(2)}(x),\gamma^1\psi^{*(1)}(x) )  
\label{C}
\\
&{\rm P}:(\psi^{(1)}(x),\psi^{(2)}(x))~\to~(i\gamma^2\psi^{(2)}(x'),-i\gamma^2\psi^{(1)}(x'))
 & x'= (x_0,x_1,-x_2)\label{P}
\\
&{\rm T}:(\psi^{(1)}(x),\psi^{(2)}(x))~\to~(\mathcal T \gamma^2\psi^{(2)}(\tilde x),\mathcal T \gamma^2\psi^{(1)}(\tilde x))
 &\tilde x=(-x_0,x_1,x_2)\label{T}
\end{align}
 
 We note that there is no valley symmetry in this model.  The interchange of  the two valleys is not
a symmetry because  the valleys have different boundary conditions.   Also,
note that the valleys are interchanged under any one of the C, P or T transformations.  This means that, if we had a single valley, rather than two valleys, none of  C, P or  would be symmetries. and CPT would be a symmetry.  This is consistent with the fact that the theory is not Lorentz invariant, since a Lorentz transformation would have to preserve the valley index, whereas CPT made from the transformations above necessarily interchanges the valleys.

\section{Explicit solution}

With the choice of Dirac gamma-matrices in equation (\ref{dirac matrices}), the explicit forms of the Dirac
 equation (\ref{de}) and boundary conditions (\ref{bc}) are
 \begin{align}
&\left[ \begin{matrix}-i\partial_0 & \partial_1-i\partial_2 \cr \partial_1+i\partial_2 & i\partial_0\cr\end{matrix}\right]
\left[ \begin{matrix}u^{(A)} (x)\cr v^{(A)}(x)\cr \end{matrix}\right]=0~,~A=1,2
\label{dirac equation} \\
&u^{(1)} (x^1\to0)=0 ~,~v^{(2)} (x^1\to0) =0
\label{bc1}\end{align}
The equation (\ref{dirac equation}) and the boundary conditions (\ref{bc1})  have the explicit solutions 
\begin{align}
&\psi^{(1)}_{\omega k\ell}(x)= 
\frac{e^{-i\omega x^0 +ik x^2}}{\sqrt{2}\pi}
\left[ \begin{matrix} \sin \ell x^1 \cr  \frac{k}{\omega}\sin \ell x^1 + \frac{\ell}{\omega}\cos \ell x^1 \cr \end{matrix}\right]
\label{bulk1}\\
&\psi^{(2)}_{\omega k\ell}(x)= 
\frac{e^{-i\omega x^0 +ik x^2}}{\sqrt{2}\pi}
\left[ \begin{matrix} \frac{k}{\omega}\sin \ell x^1 -\frac{\ell}{\omega}\cos \ell x^1 \cr \sin \ell x^1  \cr \end{matrix}\right]
\label{bulk2}\\ &\ell>0~,~~\omega=\pm\sqrt{k^2+\ell^2} \label{frequencies} \\
&\psi^{(1)}_{k}(x)= \sqrt{\frac{k}{\pi}}\left[ \begin{matrix}0\cr e^{-k(x^1-ix^2)}\cr \end{matrix}\right] ,~k>0
\label{edge1} \\ &
\psi^{(2)}_{k}(x)= \sqrt{\frac{-k}{\pi}} \left[\begin{matrix} e^{k(x^1 + ix^2)}\cr 0\cr \end{matrix}\right] ,~k<0
\label{edge2}
\end{align}

The solutions   $\psi^{(A)}_{\omega k \ell}(x)$ are Dirac spinors which propagate in the bulk of the half-space and we shall refer to them as ``bulk states''.   They have two signs of the frequency, $\omega$ and the wave-number $k$.  Their plane-wave parts $\sim e^{-i\omega x^0 +ik x^2}$ carry a representation of translations in time and along the $x^2$ direction, parallel to the edge.  

The solutions $\psi^{(A)}_{k}(x)$ which are localized near the edge of the half-plane at $x^1=0$  are referred to  as ``edge states''.  They also carry a representation of translations parallel to the edge $\sim e^{ik x^2}$, and time translation where they have zero frequency. They have the peculiar feature that their momentum quantum numbers, $k$, for edge states in a given valley have a the same sign for the entire set of edge wave-functions.

The bulk and edge states obey a completeness relation
\begin{align}
\sum_\omega\int dkd\ell \psi^{(A)}_{\omega k\ell}(x^0,x^1,x^2) \psi^{(A)\dagger}_{\omega k\ell}(x^0,\tilde x^1,\tilde x^2) 
+\int dk  \psi^{(A)}_{k}(x^1,x^2) \psi^{(A)\dagger}_{k}(\tilde x^1,\tilde x^2) \nonumber \\ 
=\delta(x^1-\tilde x^1)\delta(x^2-\tilde x^2)~,~~A=1,2
\label{completeness}
\end{align}
which can be checked explicitly and which therefore implies that we have found all of the solutions. 
  
  Given the solutions of the Dirac equation in equations (\ref{bulk1})-(\ref{edge2}), the quantized Dirac field is
\begin{align}
  &\psi^{(A)}(x) =
  \int dkdl\left\{ \psi^{(A)}_{ \sqrt{k^2+\ell^2}k\ell}(x)a^{(A)}(k,\ell)+\psi_{-\sqrt{k^2+\ell^2}-k,\ell}(x)b^{(A)\dagger}(k,\ell) \right\} + \int dk \psi^{(A)}_k(x)a^{(A)}(k) 
  \label{second quantized}
\end{align}
 Completeness of the solutions (\ref{completeness}) implies that the second quantized field in equation (\ref{second quantized}) satisfies the canonical equal-time anti-commutation relation for the Dirac field 
 \begin{align}
 &\left\{\psi^{(A)}_a(x),\psi^{(B)\dagger}_b(y)\right\}_{x^0=y^0}=\delta_{ab}\Delta^{AB}\delta^2(x-y)\\
 & \left\{\psi^{(A)}_a(x),\psi^{(B)}_b(y)\right\}_{x^0=y^0}=0
  ~,~
   \left\{\psi^{(A)\dagger}_a(x),\psi^{(B)\dagger}_b(y)\right\}_{x^0=y^0}=0
   \end{align}
 when $a^{(A)}(k,\ell)$, $b^{(A)}(k,\ell)$, $a^{(A)}(k) $ and their Hermitian conjugates have the usual anti-commutation relations of fermion and anti-fermion annihilation and creation operators, respectively. 
 
 In particular, the creation and annihilation operators for bulk and edge degrees of freedom anti-commute with each other and we can find a basis of the set of states on which they operate that is composed of direct products of bulk and edge states.
 We are interested in the states of this system near the Dirac vacuum where all of the negative energy bulk states are occupied and all of the positive bulk energy states are empty, 
 $$
 a^{(A)}(k,\ell)|0>_{\rm bulk}=0~,~~ b^{(A)}(k,\ell)|0>_{\rm bulk}=0
 $$
  This state is unambiguous for the bulk states where there is a clear distinction between positive and negative energy states.  The edge states must  carry a representation of the anti-commutator algebra of the edge state creation and annihilation operators,
 \begin{align}
& \{ a^{(A)}(k),a^{(B)\dagger}(k')\}=\delta^{AB}\delta(k-k') \label{edge state algebra1}\\
&  \{ a^{(A)}(k),a^{(B)}(k')\}=0~,~
   \{ a^{(A)\dagger}(k),a^{(B)\dagger}(k')\}=0 \label{edge state algebra2}
   \end{align}
 We will denote a generic state in a representation of the algebra in equations (\ref{edge state algebra1}) and (\ref{edge state algebra2}) by $|{\rm edge}>$ and we will call then edge states. 
 We will be interested in states of the form
 $$
 |0>_{\rm bulk}\otimes|{\rm edge}>
 $$
 In the absence of interactions of the fermions, there is an enormous degeneracy of possible edge states in that any distribution of fermions in the edge states has the same energy as any other distribution.  It is a priori unclear whether there are some edge states which are
 distinguished from others.   Understanding possible answers to this question will be our goal in the following.
 
 \section{Scale invariant states}
 
It is interesting to ask what the possible scale invariant states of this scale invariant theory might be.  Since the Dirac field has classical dimension one, a scale transformation, $x^\mu\to\Lambda x^\mu$ acts on the Dirac field by the replacement
\begin{align}\label{scale operator}
\psi^{(A)}(x) \to \Lambda\psi^{(A)}(\Lambda x)
\end{align}
The wave-functions 
for the positive, negative and edge states then transform as
\begin{align}
 & \psi^{(A)}_{\omega k \ell}(  x) ~~\to~~\Lambda\psi^{(A)} _{\omega k \ell}(\Lambda x)~=~ \Lambda\psi^{(A)}_{\Lambda\omega \Lambda k\Lambda\ell}(x)\\
&\psi ^{(A)}_{k}( x)   ~~\to~~ \Lambda\psi ^{(A)}_{k}(\Lambda x)~=~  \Lambda^{\frac{1}{2}}\psi^{(A)}_{\Lambda k}(x)
\end{align}

In order to construct scale invariant states, we are interested in complete orbits of the set of all possible scale transformations. 
An orbit of the scaling transformations acting on a function is the set of all functions that are generated from it by scale transformation with  $0<\Lambda<\infty$.
An orbit of a particular bulk state under scaling is thus the infinite set of bulk states with all magnitudes but the same signs of $k, \ell$ and $\omega$,
$$
{\rm Orbit~of~scaling}~=~\{ \Lambda\psi ^{(A)}_{\Lambda\omega \Lambda k\Lambda\ell }(x), \forall \Lambda\in (0,\infty) \}
$$
The set of all bulk states thus divides into four orbits of the scaling transformations according to the signs of $k$ and $\omega$, each subset being an orbit of scaling. (Remember that for bulk states $\ell$ is always positive.) The Dirac ground state of the bulk, with all negative $\omega$ and all positive and negative values of $k$ therefore has complete orbits of the scale transformation and it is scale invariant. Another way to put this is to say that the operator equations which define the Dirac vacuum for the bulk states 
\begin{align}
a(k,\ell) |0>_{\rm bulk}=0~,~~b(k,\ell)|0>_{\rm bulk}=0 ~,~~\forall k,\ell
\end{align}
is compatible with the existence of a unitary operator, $\Omega(\Lambda)$, 
acting on the bulk sector which
implements
\begin{align}
\Omega(\Lambda) a^{(A)}(k,\ell)\Omega^\dagger(\Lambda)
=\Lambda a^{(A)}(k/\Lambda,\ell/\Lambda)
~,~~
\Omega(\Lambda) a^{(A)\dagger}(k,\ell)\Omega^\dagger(\Lambda)
=\Lambda a^{(A)\dagger}(k/\Lambda,\ell/\Lambda)
\\
\Omega(\Lambda) b^{(A)}(k,\ell)\Omega^\dagger(\Lambda)
=\Lambda b^{(A)}(k/\Lambda,\ell/\Lambda)
~,~~
\Omega(\Lambda) b^{(A)\dagger}(k,\ell)\Omega^\dagger(\Lambda)
=\Lambda b^{(A)\dagger}(k/\Lambda,\ell/\Lambda)
\end{align}
and which leaves the bulk vacuum invariant,
\begin{align}
\Omega(\Lambda)|0>_{\rm bulk} =|0>_{\rm bulk}
\label{bulk vacuum invariant}
\end{align}
 
 On the other hand, as can be seen by inspecting equations (\ref{edge1}) and (\ref{edge2}), the set of all of the edge states in a given valley comprise a single orbit of scaling, 
\begin{align}\label{orbit of scaling}
{\rm Orbit~of~scaling}~=~\{ \Lambda^{\frac{1}{2}}\psi ^{(A)}_{\Lambda k}(x), \forall \Lambda\in (0,\infty) \}
\end{align}
The only possible scale invariant states of the edge in a particular valley must then have either all of the edge states completely filled or all of the edge states completely empty.  Amongst all of the possible fillings of the edge states for each valley, it is only these two which are scale invariant. There are thus four possibilities for a scale-invariant edge state:
\begin{align}
&a^{(1)}(k)|ff>=0~~ \forall k>0~,~~a^{(2)}(k)|ff>=0 ~~\forall k<0 
\label{pos1}  \\
&a^{(1)\dagger}(k)|ef>=0~~ \forall k>0~,~~a^{(2)}(k)|ef>=0 ~~\forall k<0 
\label{pos2} \\
&a^{(1)}(k)|fe>=0~~ \forall k>0~,~~a^{(2)\dagger}(k)|fe>=0 ~~\forall k<0 
\label{pos3} \\
&a^{(1)\dagger}(k)|ee>=0~~ \forall k>0~,~~a^{(2)\dagger}(k)|ee>=0 ~~\forall k<0 
\label{pos4} 
\end{align}
The scale invariant states of the entire system are then
\begin{align}
|0>_{\rm bulk} \otimes|ff>~,~~
|0>_{\rm bulk} \otimes|ee>~,~~
|0>_{\rm bulk} \otimes|fe>~,~~
|0>_{\rm bulk} \otimes|ef>
\end{align}

\section{Momentum density at the edge}

Let us examine the possibility that these states
could have differing physical properties. 
Each  state with wave-function in the bulk, $\psi ^{(A)}_{\omega k \ell}(x)$,  or at the edge, $\psi ^{(A)}_{k}(x)$, carries a linear momentum $k$. This momentum is directed parallel to the boundary.  Moreover, as we have already observed, all of the edge  states  in a given valley have the same sign of $k$. The contribution of a given occupied  bulk state to the total momentum is
$
k \psi ^{(A)\dagger}_{\omega k \ell }(x)\psi^{(A)}_{\omega k\ell }(x)
$
and the contribution of an occupied edge state is 
$
k\psi^{(A)\dagger}_{k}(x)\psi ^{(A)}_{k}(x)
$.

The linear momentum is the spatial volume integral of the stress tensor
\begin{align}
{\bf T}^{\mu\nu}(x)=\frac{i}{4}\sum_A \bar\psi^{(A)}(x)(\gamma^\mu\overrightarrow\partial^\nu
-\gamma^\mu\overleftarrow\partial^\nu+\gamma^\nu\overrightarrow \partial^\mu-
\gamma^\nu\overleftarrow \partial^\mu)\psi^{(A)}(x)
\end{align}
We expect that parity symmetry of the bulk states would lead to cancellation of the linear momentum.  It is instructive to confirm this more explicitly.  If we add up the contribution of the occupied bulk states  (after some algebra) we find
\begin{align}
&<{\bf T}^{02}(x)>_{\rm bulk} 
=\lim_{x'\to x}\frac{1}{i}\frac{d}{d{x}^2}\int_{-\infty}^\infty dk\int_0^\infty d\ell  \psi ^{(A)\dagger}_{-\sqrt{k^2+\ell^2}  k \ell}(x')\psi^{(A)} _{-\sqrt{k^2+\ell^2}  k \ell}(x)
\end{align}
We can use a particle-hole-symmetry (CP), which is a good symmetry of each valley separately, to re-write the 
right-hand-side of the above equations as
\begin{align}
&<{\bf T}^{02}(x)>_{\rm bulk} = \nonumber \\
&\sum_A\lim_{x'\to x}\frac{1}{2i}\frac{d}{dx^2}\int_{-\infty}^\infty d\ell \int_0^\infty dk
\left[ \psi ^{(A)\dagger}_{\sqrt{k^2+\ell^2}  k \ell}(x')\psi^{(A)} _{\sqrt{k^2+\ell^2}  k \ell}(x)
+\psi ^{(A)\dagger}_{-\sqrt{k^2+\ell^2}  k \ell}(x')\psi^{(A)} _{-\sqrt{k^2+\ell^2}  k \ell}(x)
\right]
\end{align}
Then we can use completeness of the set of all wave-functions to rewrite the right-hand-side of the above equation as
\begin{align}
&<{\bf T}^{02}(x)>_{\rm bulk} = \nonumber \\
&\sum_A\lim_{x'\to x}\frac{1}{i}\frac{d}{dx^2} \;\frac{1}{2}\biggl\{
\delta^2(x'-x)-\int dk  \psi ^{(A)\dagger}_{k}(x')\psi ^{(A)}_{k}(x) 
\biggr\}
 =-\frac{1}{2}\sum_A\int dk~k~\psi ^{(A)\dagger}_{k}(x)\psi^{(A)} _{k}(x)\,, 
\end{align}
where we have taken the $x'\to x$ limit and we have assumed that the delta function
(or  a parity symmetric regularization of the delta function) obeys $\lim_{x'\to x}\frac{d}{dx_2}\delta^2(x-x')=0$. 
Now, we note that 
\begin{align}
\int  dk  ~k~ \psi ^{(1)\dagger}_{k}(x)\psi^{(1)} _{k}(x)=-\int  dk ~ k~ \psi ^{(2)\dagger}_{k}(x)\psi^{(2)} _{k}(x)
\end{align}
Our conclusion is that the bulk Dirac sea contributes to the expectation value of the stress tensor cancel  
\begin{align}
 <{\bf T}^{02}(x)>_{\rm bulk}=0
  \end{align}
  Here, we note that this is due to an intricate cancellation between
  the momenta in each of the valleys. If there were only one valley, the Dirac sea of the bulk fermions would carry momentum density.  
  
  Now we see that momentum density distinguishes the scale invariant states of the edge.  The states 
  $|ff>$ and $|ee>$ where all of the edge states in both valleys are completely filled, or all of the edge states in both valleys are completely empty will carry no linear momentum, in that it cancels between the valley contributions.  The other two states $|ef>$ and $|fe>$ will carry a linear momentum density
  which is given by
 \begin{align}\int_0^\infty   dk  ~k~\psi ^{(1)\dagger}_{k}(x)\psi^{(1)} _{k}(x) 
 =\frac{1}{4\pi}\frac{1}{[x^1]^3}
\end{align}
We conclude that the scale invariant states have the properties
   \begin{align}
 & <ef|\otimes_{\rm bulk}<0| {\bf T}^{02}(x) |0>_{\rm bulk}\otimes|ef>
 =-\frac{1}{4\pi}\frac{1}{[x^1]^3}
\\
& <fe|\otimes_{\rm bulk}<0| {\bf T}^{02}(x) |0>_{\rm bulk}\otimes|fe>
  =\frac{1}{4\pi}\frac{1}{[x^1]^3}
\\ 
 & <ff|\otimes_{\rm bulk}<0| {\bf T}^{02}(x) |0>_{\rm bulk}\otimes|ff>
  =0
\\ \\
 & <ee|\otimes_{\rm bulk}<0| {\bf T}^{02}(x) |0>_{\rm bulk}\otimes|ee>
  =0
\end{align}
Another way to see that this should be the case is to note that P parity symmetry and translation invariance would make the above one-point functions vanish and it is only the states $ |0>_{\rm bulk}\otimes|0>_{ee}$ and $ |0>_{\rm bulk}\otimes|0>_{ff}$ which have both of these symmetries.  The states $ |0>_{\rm bulk}\otimes|0>_{fe}$ and $ |0>_{\rm bulk}\otimes|0>_{ef}$ are a doublet which transform into each other under P.

\section{U(1) charge density at the edge}

By an argument similar to the previous section where we studied the momentum density, we could examine U(1) currents.  We note that each valley has a U(1) symmetry and each should have a conserved U(1) current density. Here we will only consider 
  the total $U(1)$ current density which is the sum of the two. We define the operator current density using point splitting and we make it transform correctly under  C and P  by using the Dirac commutator operator ordering,
\begin{align}
J^{\mu}(x)\equiv\sum_A\lim_{x'\to x}[\gamma^0\gamma^\mu]_{ab} \frac{1}{2}\left[ \psi^{(A)\dagger}_a(x'),\psi^{(A)}_b(x)\right]
\end{align}
Since, under the $C$ transformation 
\begin{align}
C:~J^{\mu}(x)\to- J^{\mu}(x)
\end{align}
and since the bulk states transform covariantly under $C$, the contribution of the bulk states to the
expectation value 
$$
<{\rm edge}|\otimes_{\rm bulk} <0|~J^{\mu}(x) |0>_{\rm bulk}\otimes|{\rm edge}>
$$
must vanish and the entire contribution comes from the edge states. The 
edge state wave-functions obey
\begin{align}
\int_0^\infty dk \psi^{(1)\dagger}_k(x)\psi^{(1)}_k(x)=\int_{-\infty}^0 dk \psi^{(2)\dagger}_k(x)\psi^{(2)}_k(x)=
\frac{1}{4\pi}\frac{1}{[x^1]^2}
\end{align}
Then, adding the edge state contribution to the vanishing bulk contribution yields the U(1) current density, for example, for the completely filled edge states
\begin{align}
&<ff|\otimes_{\rm bulk} <0|~J^{0}(x) |0>_{\rm bulk}\otimes|ff>=
\frac{1}{2}\left[\int_0^\infty dk \psi^{(1)\dagger}(x)\psi^{(1)}(x)+\int_{-\infty}^0 dk \psi^{(2)\dagger}(x)\psi^{(2)}(x)\right]
\end{align}
The factor of $\frac{1}{2}$ on the right-hand-side comes from the Dirac commutator. 
This leads to the one-point functions for the U(1) current in each of the scale invariant states,
\begin{align}
&<ff|\otimes_{\rm bulk} <0|~J^{\mu}(x) |0>_{\rm bulk}\otimes|ff>=\frac{\delta^\mu_0}{4\pi}\frac{1}{[x^1]^2}
\\
&<ee|\otimes_{\rm bulk} <0|~J^{\mu}(x) |0>_{\rm bulk}\otimes|ee>=-\frac{\delta^\mu_0}{4\pi}\frac{1}{[x^1]^2}
\\
&<ef|\otimes_{\rm bulk} <0|~J^{\mu}(x) |0>_{\rm bulk}\otimes|ef>=0
\\
&<fe|\otimes_{\rm bulk} <0|~J^{\mu}(x) |0>_{\rm bulk}\otimes|fe>=0
\end{align}
In the previous section we concluded that the P invariant states $ |0>_{\rm bulk}\otimes|ff>$ and $ |0>_{\rm bulk}\otimes|ee>$ had vanishing momenta.  However, these states are not C invariant. They form a doublet under C and, as a consequence, they have a nonzero U(1) charge densities, with opposite signs. 

\section{Epilogue}

In conclusion, we have outlined some properties of the scale invariant states of a simple scale invariant but not Lorentz or conformal invariant boundary field theory. The existence of edge states leads to  the interesting anomalous effects that we find.  For our model with two valleys, there are four scale invariant states which split into two pairs.   One pair of scale invariant edge states  are seperately P parity invariant but they form a doublet of  C charge conjugation symmetry. The other pair  are separately C invariant but form a doublet of  P. The states in the C doublet  have oppositely signed nonzero charge densities whereas the states in the P doublet have nonzero oppositely signed linear momentum densities. The latter linear momentum is directed parallel to the edge and the charge and momentum densities in both cases decay by a scale coraviant power law with distance from the edge
and with an oveall coefficient which is independent of the details of the system. 

Edge states of the kind that we have studied here has long been known to occur at the lattice level in, for example, the tight binding model of graphene electrons and the have long been conjectured to have interesting properties and potentially important physical applications \cite{Fukita}
\cite{Nakada}. 
 For example it is known that states very similar to those which we have considered here but in a theory with an additional spin degree of freedom have anti-ferromagnetic states and there is a proof that they are the ground states of a sufficiently weakly coupled class of two-body interactions that could be added to the free Dirac theory are indeed ferromagnetic \cite{Semenoff:2022azt} and the ferromagnetic states are direct analogs of the scale invariant states that we have discussed here. It would be very interesting to study interacting field theories with the boundary and boundary conditions that we have discussed.
 
This work is supported in part by the Natural Sciences and Engineering Research Council of Canada (NSERC).


\begin{thebibliography}{99}


\bibitem{Semenoff:1984dq}
G.~W.~Semenoff,
``Condensed Matter Simulation of a Three-dimensional Anomaly,''
Phys. Rev. Lett. \textbf{53}, 2449 (1984)

\bibitem{CastroNeto:2007fxn}
A.~H.~Castro Neto, F.~Guinea, N.~M.~R.~Peres, K.~S.~Novoselov and A.~K.~Geim,
``The electronic properties of graphene,''
Rev. Mod. Phys. \textbf{81}, 109-162 (2009)
[arXiv:0709.1163 [cond-mat.other]].

\bibitem{ak1}A.R.~Akhmerov, C.W.J.~Beenakker, ``Boundary conditions for dirac fermions on a terminated honeycomb lattice'', Phys. Rev. \textbf{B 77} (2008) 085423.

\bibitem{ak2}J.A.M~ van Ostaay, A.R.~Akhmerov, C.W.J.~Beenakker, M.~Wimmer, ``Dirac boundary condition at the reconstructed zigzag edge of graphene'', Phys. Rev.\textbf{ B 84} (2011) 195434.

\bibitem{Biswas:2022dkg}
S.~Biswas and G.~W.~Semenoff,
``Massless fermions on a half-space: the curious case of 2+1-dimensions,''
JHEP \textbf{10}, 045 (2022)
[arXiv:2208.06374 [hep-th]].

\bibitem{Fukita}M.~Fujita, K.~Wakabayashi, K.~Nakada,~K. Kusakabe, ``Peculiar Localized State at Zigzag Graphite Edge'', J. Phys. Soc. Jpn. \textbf{65},1920v (1996).

\bibitem{Nakada}K.~Nakada, M.~Fujita, G.~Dresselhaus, M.S.~Dresselhaus, ``Edge state in graphene ribbons: Nanometer size effect and edge shape dependence'', Phys. Rev. \textbf{B 54}, 17954 (1996).


\bibitem{Semenoff:2022azt}
G.~W.~Semenoff,
``Boundary ferromagnetism in zigzag edged graphene,''
J. Math. Phys. \textbf{64}, no.7, 071902 (2023)
[arXiv:2211.09282 [cond-mat.mes-hall]].




\end{thebibliography}
\end{document}